\documentclass{vgtc}                          

\graphicspath{{figures/}{pictures/}{images/}{./}} 

\usepackage{times}                     

\usepackage{tabu}                      
\usepackage{booktabs}                  
\usepackage{lipsum}                    
\usepackage{mwe}                       

\usepackage{mathptmx}                  

\onlineid{1040}

\vgtccategory{Research}

\vgtcinsertpkg

\title{Data Guards: Challenges and Solutions for Fostering Trust in Data}

\author{Nicole Sultanum\thanks{e-mail: nsultanum@tableau.com}\\ %
        \scriptsize Tableau Research, Seattle, WA, USA %
\and Dennis Bromley\thanks{e-mail: dbromley@tableau.com}\\ %
     \scriptsize Tableau Research, Seattle, WA, USA %
\and Michael Correll\thanks{e-mail: m.correll@northeastern.edu}\\ %
     \scriptsize Northeastern University, Portland, ME, USA}

\teaser{
  \centering
  \includegraphics[alt={Seven idea cards communicating our data guards, distributed across three groups: Overview, Details, and Community. There are three data guards idea under the Overview group. 1. Data and Pipeline tests. Pre-defined assertions about the data pipeline or the data itself that reflect data quality issues. Could be user-created, and could have alerts.
For example: a, presence of null values; b, percent values add up to 100\%; c,
data rows are not being subsampled. 2. Data Quality Agent. Automated analysis entity that identifies anomalies or other problems in the data (such as null values) that may be worth investigating. 3. Data and Pipeline Update Alerts. A notification appears when something in the data or the data pipeline has been updated. There are two ideas under the Details group. 1. Explanation and Status. Written, and potentially visual, report or narrative that describes the data and the data processing pipeline in lay terms. Could be manually created, or automatically generated and updated to reflect changes in the pipeline. 2. Data Traces. A high level, potentially visual and narrative overview of the pipeline that can be interrogated. Could be a general overview of the whole dashboard, or focused on segment of data. Can be drilled down; an expert user could walk viewers through if more detail is desired. Finally, there are two ideas under the community group. 1. Stamp of Approval. One or more trusted and relevant people (e.g., data steward, specialist, etc) approved the data source, pipeline and/or dashboard display. An official record of human oversight. 2. Crowd Wisdom. Provide viewers a feedback channel for discussion, consensus and evaluation. For example, viewers can flag issues on a dashboard element, share their personal takeaways from the data, and discuss data findings.},width=\textwidth]{dg-camera-ready-idea-cards.pdf}
  
  \vspace{-0.5em}
  \caption{Proposed solutions for establishing trust in data, grouped into three categories: \textcolor{overview}{\textbf{Overview}}, \textcolor{details}{\textbf{Details}}, and \textcolor{community}{\textbf{Community}}. Solutions were derived from a series of data-producer interviews based upon interviewee's current data-trust obstacles and opportunities, and validated by data \textit{consumers} in a card sorting exercise (Figure \ref{fig:card_sorting_tally}). Card design used in studies is illustrated above.}
  \label{fig:teaser}
}

\abstract{

From dirty data to intentional deception, there are many threats to the validity of data-driven decisions. Making use of data, especially new or unfamiliar data, therefore requires a degree of trust or verification. How is this trust established? In this paper, we present the results of a series of interviews with both producers and consumers of \textit{data artifacts} (outputs of data ecosystems like spreadsheets, charts, and dashboards) aimed at understanding strategies and obstacles to building trust in data. We find a recurring need, but lack of existing standards, for data validation and verification, especially among data consumers. We therefore propose a set of \textit{data guards}: methods and tools for fostering trust in data artifacts.
} 

\usepackage{soul}
\usepackage{fontawesome5}
\usepackage{comment}

\definecolor{overview}{RGB}{89,138,51}
\definecolor{details}{RGB}{46,96,182}
\definecolor{community}{RGB}{189,53,39}

\begin{document}


\firstsection{Introduction}

\maketitle

Errors in data-driven decision-making can have drastic consequences, from the copy and paste errors that some suggest led to over 100,000 excess deaths~\cite{herndon2014does,watkins2017effects} to the intentional deception that produced high-profile scandals in biology and psychology~\cite{lange_ferret_2023}. 
Despite the risks, people at all levels of data work must continue to make data-driven decisions in both personal and professional capacities. But people need to \textit{trust} their data in order make decisions. How does one build trust, especially in new or unfamiliar data prepared or curated by others? 

Prior work suggests that strategies for building trust in data are often informal (such as ``eyeballing'' of tabular data~\cite{bartram2021untidy}), often undocumented~\cite{ruddle_tasks_2024}, and that the resulting trustworthiness of data artifacts like visualizations can be difficult to measure~\cite{elhamdadi2022we}. Existing tooling for validation and verification are often targeted towards users who are creating or manipulating datasets~\cite{kandel2012profiler}, rather than the broader category of what prior work has called \textit{data workers}: those who regularly work with data but do not identify as data scientists~\cite{liu2019understanding,tory2021finding}. These data workers may in turn be \textit{consumers} of data artifacts: asked to trust in data they did not produce~\cite{tory2021finding}.

In this work, we contribute findings from a series of interviews to understand the struggles and strategies of data workers when vetting or validating \textit{data artifacts} (\S\ref{sec:interviews}).
Participants report frustration with current tooling to support these strategies, from 
the inability to verify high-level assumptions 
to the dependence on third parties to diagnose issues.
Based on these findings and inspired by work in data transparency, we propose a set of consumer-centered proto-solutions we term \textit{data guards}: ways of fostering trust or transparency in data communication that are easy for \textit{producers} (i.e., data artifact creators) to author, and for \textit{consumers} to understand (\S\ref{sec:dataguards}). 
We end with a call to action for the visualization community to more tightly integrate data guards into the design, evaluation, and communication of data artifacts (\S\ref{sec:discussion}).

\section{Related Work}

Our work focuses on how \textit{consumers} of data artifacts build trust, validate, or otherwise confirm insights they glean from data, and strategies that \textit{producers} of these artifacts can employ to assist in or otherwise augment these trust-building processes. While there is extensive existing literature on data \textit{cleaning}~\cite{chu2016data,kandel2012enterprise,liu2018steering}, data \textit{profiling}~\cite{naumann2014data}, data \textit{curation}~\cite{muller2019data}, and otherwise avoiding \textit{dirty} data~\cite{kim2003taxonomy} for those creating datasets, we focus on both different audiences (those who make use of these resulting ``clean'' datasets, who may or may not be the same populations as those who created them), and different tasks (confirming that a finding from, e.g., a table or dashboard, is sufficiently reliable).

We focus on three areas of prior work: analyses of work practices and sensemaking among data workers---particularly data consumers; analyses of errors, omissions, and verification strategies in data-driven decision-making;
and tooling and strategies for verification and validation in analytics.

\subsection{Data Workers and Trust-Building}
Prior works on trust and verification, such as 
Kandel et al.~\cite{kandel2012enterprise} and surveys of data quality management such as Liu et al.~\cite{liu2018steering}, often focus on data scientists charged with the \textit{creation} of high-quality datasets. 
We instead shift our attention to a particular subset of ``data workers''~\cite{bartram2021untidy,crisan2020passing,tory2021finding} that we call \textit{consumers}: that is, data workers who make use of data artifacts created by others.

In a study on the related process of data profiling by analysts, Ruddle et al.~\cite{ruddle_tasks_2024} notably claim that ``most analysts perform profiling in an ad-hoc manner, following an undocumented process that makes data
profiling more an art than a science.'' This finding aligns with Bartram et al.'s~\cite{bartram2021untidy} survey of data munging and federation practices of data workers, which found that qualitative processes like ``eyeballing'' were prominent in building trust in data, often in conjunction with extensive use of spreadsheet tools as ways to provide a sense of direct control or ownership over their data. The participants in the more general analysis of analytic processes in Kandogan et al.~\cite{kandogan2014data} similarly report that ``that 80 percent of any
effort is getting the data together and understanding data quality'', but that challenges of provenance and communication can result in data quality concerns that require ``checks and balances'' to detect and correct. Of note is that data quality issues do not only arise either in the creation of datasets or in the interpretation of final visualizations, but rather that, as per Sacha et al.~\cite{sacha_role_2016}, uncertainties occur at all stages of the analytics process, with impacts that flow between stages. Chu et al.~\cite{chu2016data} further emphasize that data cleaning and data verification is complex and generally qualitative.

Beyond the steps involved in verification or trust-building, the very notion of what is meant by data quality and trustworthiness is multifaceted. For instance, in a survey of a similar participant pool as our own (viz., ``those who use data''), Wang and Strong~\cite{wang1996beyond} identify 20 dimensions under 4 categories that their participants associated with data quality, several which touch on aspects of trust: accuracy of data (e.g., believability, accuracy, reputation), relevancy of data (e.g. timeliness, value-add), representation of data (e.g., interpretability, ease of understanding), and accessibility of data.

\subsection{Smells and Mirages: Threats to Data Reliability}

While we acknowledge 
lack of trust arising from intentional deception~\cite{lisnic2023misleading}, 
unintentional trust issues can arise through an unreliable finding or a misinterpretation of a visualization. McNutt et al.~\cite{mcnutt2020surfacing} introduce the term ``visualization mirage'' to refer to erroneous conclusions emerging from cursory readings of a visualization without additional scrutiny. 
In line with our notion of ``data guards'', McNutt et al. propose testing regimes in order to surface potential data quality or uncertainty issues in visualizations.
Drawing from the software engineering metaphor of ``code smells'' (patterns in code that often indicate potential issues), Shome et al.~\cite{shome2022data} and Foidl et al.~\cite{foidl2022data} propose categorizations for ``data smells'', 
that can point to potential issues with data quality and interpretation. Kim et al.~\cite{kim2003taxonomy} also propose a ``taxonomy of dirty data'' to illustrate ways that data (especially manually entered or curated data) can be invalid.

Many existing visual analytics systems lack tools or workflows for detecting potentially unreliable or erroneous conclusions. This gap can result in high error rates when assessing the accuracy of insights from visual analytics~\cite{zgraggen_investigating_2018}, rates that are only partially improved by including uncertainty information~\cite{sarma_odds_2024}. In keeping with the highly contextual nature of assessing data quality, we note work by Song et al.~\cite{song2018s} suggesting the ways in which data quality information is visualized (e.g., whether missing values are elided or indicated with imputed values) can influence the perceived reliability of data sets. The inclusion of explicit visualizations of missing data can even impact the strategies that analysts use when making sense of data~\cite{song2021understanding}. The heterogeneous nature of analytical strategies can, however, exacerbate issues caused by a lack of verification in analytics. For instance, Yanai and Lercher~\cite{yanai_hypothesis_2020} find that participants who approach a dataset with a specific hypothesis in mind were less likely to notice dramatic data quality issues.

\subsection{Instrumenting Trust: Profiling and Verification}

Given the prominence of data quality concerns in analytics but the lack of standard repeatable and quantitative methods for verifying data quality, there have been a variety of proposed design interventions to assist users in either profiling, wrangling, or verifying data. 
Kandel et al.'s~\cite{kandel2012profiler} \textit{Profiler} visualizes potential anomalies in data that may indicate underling data quality issues or errors in reshaping; Jannah's~\cite{jannah2014metareader} \textit{MetaReader} surfaces errors and warnings such as potential mismatches between data columns and inferred semantic roles (e.g., a primary key);
Bors et al.'s~\cite{bors2018visual} \textit{MetricDoc} surfaces data quality metrics at both global and individual value levels (e.g., a ``plausibility'' value flagging potential status as an outlier or data quality concern); and the \textit{Ferret} system by Lange et al.~\cite{lange_ferret_2023} surfaces features that might indicate manipulation in tabular data.
However, these systems are mostly designed for those \textit{creating} data artifacts. 
Relatively fewer works focus on consumers of data artifacts, who have domain expertise but not necessarily data processing expertise or data curation access~\cite{tory2021finding}. Notable exceptions include the metric-based ``risk gauges''~\cite{zhao2017safe} or quality scores~\cite{bors2018visual},
the ``metamorphic testing'' approach in McNutt et al.~\cite{mcnutt2020surfacing} to highlight the robustness of visual patterns in charts, and 
as in Fan et al.~\cite{fan2022annotating} to 
detect
potentially deceptive visualization designs. 
We draw inspiration from these similarly motivated, but conceptually distinct, design interventions when proposing our own data guards.

\section{Characterizing Data Trust Needs}
\label{sec:interviews}

We conducted two rounds of interviews. In the first round, we interviewed eight expert dashboard \textit{producers} (P01-P08), i.e., professionals who create dashboards for corporate and general public consumption, to understand how they address trust in their dashboard designs and what challenges they (and their users) face in the process (\S\ref{sec:barriers} and \S\ref{sec:findings-desiderata}).
Using thematic analysis~\cite{braun2012thematic}, we organized this feedback into themes that we use to propose seven classes of design solutions, which we collectively call \textit{data guards} (\S\ref{sec:dataguards-ideas}).
We followed up with a second round of interviews with ten professional data \textit{consumers} (C01-C10) 
to solicit their trust-related experiences when perusing data artifacts for decision making (e.g., dashboards and other forms of dynamic data reports), which helped substantiate challenges  (\S\ref{sec:barriers}) and  desiderata~(\S\ref{sec:findings-desiderata}) identified in the first round. We also performed a card-sorting exercise with these participants to encourage ideation and prioritization around our seven proposed data guards (\S\ref{sec:dataguards-validation}). Study instruments can be found in Supplemental Materials.

\subsection{Barriers for Data Trust}
\label{sec:barriers}

We first present six categories of \textit{barriers} to trust in data (B1-B6) that emerged from our interviews. That is, challenges or contextual factors that complicate the process of building trust.


\textbf{(B1) Data is heavily context-dependent.} 
A challenge reported by producers working as external consultants\,---\,but also a natural consequence of the division of labor between producers and consumers\,---\,is a lack of shared domain knowledge~\cite{muller2019data} and common ground~\cite{olson2000distance}. Lack of familiarity with domain idiosyncrasies such as how to calculate bond rate denominators (e.g., \textit{360 days? Exact days in year, including leap days? Days excluding weekends?} (P03)) can result in errors that would not be caught be generalizable data quality metrics and is important for validation. Yet, domain information is difficult to transfer:  \textit{``Sometimes the people who are building these dashboards or building these data sources don't really know the business as intimately. There's definitely a disconnect and there's just a bit of education that needs to happen.''} (C05).



\textbf{(B2) Detecting issues requires a discerning eye}. 
The opacity of data processing pipelines in data artifacts like dashboards means that consumers might only notice issues when they see something that defies their expectations (e.g., outliers). This was often reported as an intuition based on experience rather than a systematic approach, akin to ``data smells''~\cite{shome2022data, foidl2022data}: \textit{``this doesn't smell right''} (C08). Novice consumers are therefore less likely to detect issues when they arise. If discrepancies are not visible or data not inspectable, trust can be a (perhaps unjustified) default: \textit{``I do trust the data because I don't have a choice. is that fair? (..) Maybe just because  there's no other way to get that [data], I can't know the answer, I can't verify it.''} (C02).  


\textbf{(B3) Data trust builds on human relationships.} 
Data work is collaborative. Most producers (5/8) stated working closely with consumers (and proxies) to not only bridge the domain knowledge gap, but also build interpersonal trust; e.g., \textit{``They trust me because it's me. And that's the professionalism you build up (..). I've spent the time to reconcile, the data, I've thought about this and the 10 different ways that you're going to use it.''} (P07).
Consumers also state relying on interpersonal trust as part of their assessment of data trust~\cite{tory2021finding}, e.g., \textit{``If I was looking at a [data source], and they said [anonymous] certified this I'd be cool. I trust it. Because he knows what he's talking about.''} (C02); or a community assessment, e.g., \textit{``What gives me confidence is the lack of user feedback that something is wrong.”} (C10). 
However, when no such relationship exists, trust may be negatively affected, e.g.,  \textit{``Who's presenting it? Has this person credibly done this before?''} (C09).

\textbf{(B4) Trust is hard to build, easy to lose.} 
Beyond human relationships, the ongoing experience consumers have with their data artifacts also influences their perception of trust: \textit{``Some of it comes down to day-to-day working in these tools (..), and feeling like it's always a consistent experience to use it and that I see the same things each time.''} (C08). Conversely, a lack of experience may induce mistrust: e.g.,  \textit{``I have definitely doubted newer dashboards that have come out''} (C07); as well as those facing constant issues with particular sources, and having mistrust as a starting point: \textit{``I wish that my mental model was trust but verify. It is not. It is distrust until I can prove it.''} (C02). 

\textbf{(B5) Data definitions are often unclear or ambiguous.}
Consumers commented that understanding calculations and metrics at a deeper level was a constant challenge and cited various sources of ambiguity, including duplicate or scattered data sources, e.g., \textit{``So I look at reports in [data source 1] and [data source 2]. I do my forecasting on a sheet. And then I also look at the [data source 3]. There is not one single source of truth or one number that I can depend on''} (C05); but also conceptual alignment issues, with different stakeholders interpreting the same data differently: e.g, 
\textit{``Even within departments, we're talking different languages.`Attrition' to one person may mean something different; to me it means numbers, while for Human Resources it is attrition of people''}(C05).
Consumers often cited a lack of transparency on relevant metadata and documentation to help clarify these ambiguities:  \textit{``I could know what made the number if somebody wrote a definition down''} (C02). 




\textbf{(B6) Environments change and processes break.} 
Changes in the data and pipeline introduce risk and can have a significant impact on trust: \textit{``If something made it into production that shouldn't, it was done when the data updated''} (P06). They explain: \textit{``There have been cases where there were assumptions made in a workbook that might have no longer been true or we didn't take into account things that came in later on down the line.''} (P06). From a consumer's perspective, there are also contextual changes: \textit{``Where something changes in an upstream system (..), like business logic. It's very hard, the business users don't necessarily tell the people in charge of the data pipes that this changed happened, or they don't realize that this change happened and therefore suddenly the values are weird''} (C10). As in \textbf{(B5)}, a lack of documentation is a primary concern, but also a lack of change tracking or alerts.



\subsection{Mitigation Strategies and Desiderata}
\label{sec:findings-desiderata}

We collated strategies that our participants followed to establish or repair trust in data artifacts, plus ideas for improvement. We structure these themes as design goals \textbf{(G1-G5)}, each associated with one or more corresponding barriers they are meant to overcome.


\textbf{(G1) Understand domain knowledge (B1, B2, B5).}
Producers work closely with consumers when building data artifacts. Given the demands on domain knowledge, some also argue it is a \textit{``[consumer] job as the one who owns the data''} (P05) to validate data artifacts, and that consumers should be given better tools to assess issues e.g., \textit{``put the stop sign that says `you got to look at this'. We can show you where the numbers come from, but we can't look at it for you.''} (P01). 
Conversely, consumers also desired tools to help them navigate contextual ambiguity of data artifacts: \textit{``It would be lovely to be able to say `I see 20 different calculations called engagement. These are all the different ways that engagement is being calculated'.''}(C06). All these point to a need for domain-aware tools that can better support collaboration and error diagnosis.

\textbf{(G2) Provide provenance and metadata (B2, B4,~B5).} %
Consumers commented wanting more metadata: e.g., \textit{``I want a data definition. Where did this come from? Did it come from [my expected source]? Did it come from somewhere else?''} (C02). This is sometimes addressed in dashboards as \textit{``information icons (..) with a tool tip on to explain some of the metrics and (..) say, `Hey, if you're struggling with these metrics, then here's what they do, and how they work'\,''} (P04), but it is not always present.
Alternatively, some consumers also reported going back to the source data to diagnose issues: \textit{``I go in the system and I look at the raw data. I need to understand if there's a definition problem, or a logic problem. (..)''} (C02). This suggests consumers may benefit from some form of data source and pipeline description in a suitable format and appropriate level of detail (e.g., in situ ``metadata blocks''~\cite{bach2022dashboard} or even peripheral ``datasheets''~\cite{gebru2021datasheets}).

\textbf{(G3) Communicate high-level trust signals (B1, B2, B3).} %
While comprehensive details are useful for building initial trust and ``chasing smells'', they are not always needed to maintain existing trust. For example, with data quality controls and trusted oversight in place, having a \textit{``certified''} stamp is helpful: \textit{``When there is investment in a data analyst core and data governance, that's been vetted and approved everywhere it needs to be approved, then I want to trust it''} (C07). Other visual signals like breadcrumbs for filtering operations may also support data awareness that helps build trust, so \textit{``you're seeing what actually that view contains''} (P07); or warnings that something is \textit{``in a draft state''} (P04). 



\textbf{(G4) Highlight data and pipeline changes (B4, B5, B6).}
Change awareness is key. A standard practice is to communicate when data was last updated, e.g., \textit{``Everything is updated at a different cadence, and so in the dashboard, we literally have a footer under every KPI `last update by'~''} (C10). This informs that there was a change but not what has changed, leading consumers to seek assistance: \textit{``Sometimes I'll have that conversation with whoever made the updates to know exactly what they updated''} (P06). Consumers might also keep references to past data and pipeline versions (akin to version control): \textit{``I will always take snapshots of the report that I culled my numbers from. So I can link it back to, `well on this day and time this is what the report said'~''} (C05). Arguably, automation could support or supplant some of these manual strategies: \textit{``it would be cool if [my dashboard] could recognize that something funky happened between yesterday and today and flag that''} (C01)




\textbf{(G5) Leverage and mediate social trust (B1, B3).} 
Peer relationships play an important role in the data ecosystem as a primary fallback when trust collapses: \textit{``If I don't [have data governance], I'm gonna ask people I trust''} (C02). Individuals in trusted roles are also sought after to troubleshoot issues, e.g., \textit{``I actually looked for the person who created the dashboard in order to get me that information''} (C03). Peer activity also provides trust signals to consumers: \textit{``you've got this dashboard that has two views in the past six months [that should not be trusted]; and there's this one dashboard that is being only used once a year, but it's by [the CEO]''} (C03). Ultimately, data intelligence and automation are unlikely to replace trusted human relationships. However, there is little technology support to mediate these relationships currently, and the design space of social and collaborative aspects of data trust remains largely unexplored; e.g., \textit{``I don't want to have to find the person who can answer the `why'. I want this `why' to be automatically redirected to the person who could answer me''} (C03).
\begin{figure}
    \centering
    \includegraphics[alt={Table showcasing how each participant ranked each data guard. The Data Guards are sorted as follows: 1, Stamp of Approval. 2, Explanation and Status. 3, Data Traces. 4, Data Quality Agent. 5, Data and pipeline Tests. 6, Crowd Wisdom. 7, Data and pipeline Update Alerts.},width=1\columnwidth]{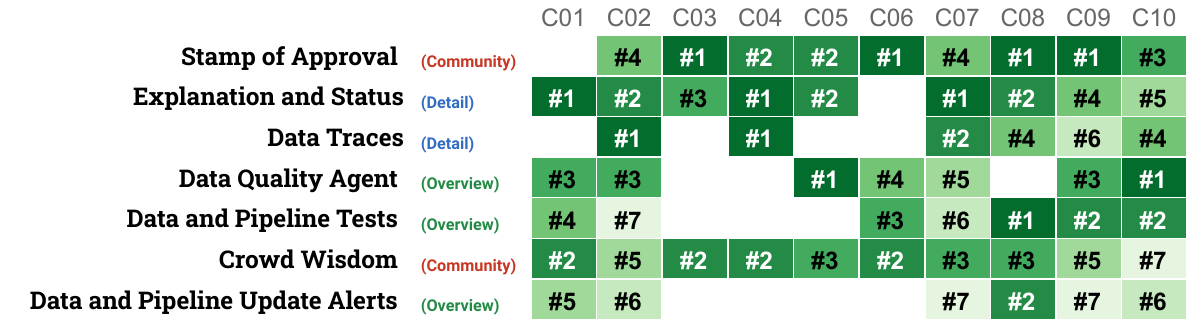}
    \caption{Card sorting results for different data-trust solutions broken out by participant and solution and colored by rank (from \#1 to \#7).  Individual boxes indicate which rank was selected by whom for each solution. Solutions are sorted vertically by recursive ranking: highest number of rank \#1's first, followed by rank \#2's, and so on.}
    \label{fig:card_sorting_tally}
    \vspace{-1em}
\end{figure}

\section{Data Guards}
\label{sec:dataguards}

We use the strategies we observed and the gaps in existing processes to propose seven classes of design interventions that we collectively call \textit{data guards} (Fig.~\ref{fig:teaser}). These solution ideas were designed as potential trust-enabling additions to data artifacts, with some inspired by current mitigation strategies and some proposing new features.
We explain our data guards alongside relevant design goals (\S\ref{sec:dataguards-ideas}) and share consumer feedback on these ideas (\S\ref{sec:dataguards-validation}).

\subsection{Data Guards Ideation}
\label{sec:dataguards-ideas}
Our seven strategies fall into three general clusters (\autoref{fig:teaser}). \textcolor{overview}{\textbf{Overview}} strategies provide high-level trust signals for monitoring (1-3). \textcolor{details}{\textbf{Details}} strategies allow for deep data dives (4-5). And, lastly, \textcolor{community}{\textbf{Community}} strategies leverage trusted peer relationships (6-7):

1. \textbf{\textcolor{overview}{\textit{Data and Pipeline Tests}} (G1, G3, G4).} Inspired by the notion of software assertions and data profiling checklists~\cite{ruddle_tasks_2024}, these strategies outline predefined tests over the data, the data pipeline, and the charts to surface data quality and presentation issues, such as null values, percentages not amounting to 100\%, truncated y-axis, or values outside domain thresholds. Rules could potentially be crafted by consumers and reused across data artifacts.

2. \textbf{\textcolor{overview}{\textit{Data Quality Agent}} (G1, G3).} 
Building off of existing data quality~\cite{bors2018visual} or insight reliability~\cite{zhao2017safe} metrics, these strategies entail automated analyses to detect ``data smells'', thus pointing to potential anomalies or other problems in the data (such as outliers or redundant variables) that consumers may find worth investigating.

3. \textbf{\textcolor{overview}{\textit{Data and Pipeline Change Alerts}} (G3, G5).}
These strategies employ notifications to communicate and summarize changes to the data or the data pipeline since the last time the user interacted with the data artifact. These summaries could augment existing metrics or notifications of data ``freshness''~\cite{bouzeghoub2004framework}.

4. \textbf{\textcolor{details}{\textit{Explanation and Status}} (G1, G2, G4).}
Drawing from existing efforts to better describe and contextualize data in machine learning, such as Model Cards~\cite{mitchell2019model}, Data Cards~\cite{pushkarna2022data} and Data Sheets~\cite{gebru2021datasheets}, these strategies employ written, potentially visual, reports describing underlying data and calculations in lay terms. These explanations could be manually created by producers, or automatically generated and updated to reflect changes in the pipeline.

5. \textbf{\textcolor{details}{\textit{Data Traces}} (G1, G2).}
These strategies provide a drill-down view of the data pipeline (potentially in narrative form) focusing on a visible data slice, e.g., from selected points in a chart, or from a data filter. Visualizations of data flows and provenance in tools like VisTrails~\cite{callahan2006vistrails} have shown promise for teaching and communication of unfamiliar visualizations~\cite{silva2011using}. These interventions could also be smaller scale, such as communicating the specific form of aggregation used to generate a mark in a visualization~\cite{kim2019designing}.

6. \textbf{\textcolor{community}{\textit{Stamp of Approval}} (G3, G5).}
These strategies rely on an explicit record of human oversight. For instance, an official record that relevant experts approved the data source, pipeline and views in this data artifact and a means to reach out to these experts.

7. \textbf{\textcolor{community}{\textit{Crowd wisdom}} (G5).}
While the stamp of approval strategy above uses the metaphor of vetting, crowd wisdom makes a less prescriptive use of a shared community shared for discussion and consensus. Consumers could flag issues, share their personal takeaways from the data, and discuss data findings (as with collaborative visualization environments like ManyEyes~\cite{viegas2007manyeyes}). Other forms of leverage of social data could be the use of records of interaction~\cite{feng2016hindsight} to guide users to important areas of a visualization, or externalizing other people's expectations of relationships in data~\cite{kim2017data}.

\subsection{Consumer Feedback on Data Guards}
\label{sec:dataguards-validation}

Consumer rankings for the data guards are shown in Fig.~\ref{fig:card_sorting_tally}. Due to time constraints, some participants chose not to rank certain items, hence the white spaces; some also chose to place two items in the same rank, hence the repeated rankings.
In this context, the item most frequently ranked first was \textcolor{community}{\textit{\textbf{Stamp of Approval}}} (4 times). Consumers who already have similar experience working with ``certified'' data sources appreciate the simplicity of having a human-vetted high-level trust signal that did not require additional investigation or action when \textit{``at that uber level where I really don't touch data''} (C09). Similarly, \textit{``Can I find the data? Can I find the right dashboard? Can I understand the dashboard? And if not, who do I contact? Realistically, given the amount of time, that would be the extent of what I could do.''} (C03).

The next most commonly top-ranked items were \textcolor{details}{\textit{\textbf{Explanation and Status}}} (3 times) and \textcolor{details}{\textbf{\textit{Data Traces}}} (2 times). They provide more in-depth trust-related information at the expense of more time and effort, aligning with past findings on data worker practices~\cite{bartram2021untidy, tory2021finding}.  Consumers argued these strategies support both building and upholding trust: \textit{``For Explanation and Status, if you don't use a dashboard for a while, if you're new to a dashboard, if you're just finding out about it if you're nearly hired, the first question you're going to ask is `how do we get to these numbers?'\,''} (C07); and \textit{``when I find something to be untrustworthy, and I then need to either convince myself it should be trusted or prove to somebody that it is not accurate, the Data Trace tells me how to do that.''} (C02). 

While these strategies in the \textcolor{details}{\textbf{Details}} cluster feature closer to the top, findings show that all proposed data guards were deemed valuable to at least some consumers. All strategies were placed on Top two at least once, and all three clusters have items ranked first. This suggests that consumer trust needs are diverse and that a multi-pronged approach to fostering trust may be most helpful.



\section{Discussion and Closing Thoughts}
\label{sec:discussion}
Trust in data artifacts is a multifaceted and nebulous concept. Beyond the issues of dirty data or intentional deception, trust in data can be lost as data circulates to new audiences, new artifacts are built on top of old data, or institutional needs or processes change. 
In general, we see a need for ways to provide context, awareness, and guidance over data artifacts, especially for \textit{consumers} of data artifacts who may have limited access to underlying data collection and curation processes that generated the artifact, who may be encountering the artifact for the first time, or, even, who may have additional domain knowledge and context that suggest threats to the validity or applicability of the data they are given.

Data guards are broad categories of designs to address these unmet trustbuilding needs. While aspects of these strategies have been proposed in prior work, our work reframes and prioritizes them with consumers in mind.
Future research is required to further explore how these strategies can be instantiated into analytics tools, and in particular the promises or limitations of automation~\cite{heer2019agency} to supplement the current highly manual and qualitative processes in trust.

These strategies can come at a cost. Data guards are themselves \textit{data artifacts}, which require the same sort of trustbuilding and validation to show their validity. Too much complexity could overwhelm users, and spurious or invalid alerts could erode trust, just as spurious or invalid data underminded our participants' trust in the artifacts they encountered.

\bibliographystyle{abbrv-doi}

\bibliography{references}
\end{document}